\documentclass[%
 reprint,
 superscriptaddress,
 amsmath,amssymb,
 aps,
pre,
floatfix,
]{revtex4-2}

\usepackage{graphicx}
\usepackage{dcolumn}
\usepackage{xcolor}

\begin{document}


\title{Temporal interaction and its role in the evolution of cooperation}

\author{Yujie He}
\affiliation{Institute of Development, Guizhou Academy of Governance, Guiyang, China}
\author{Tianyu Ren}%
 \email{tianyu.ren@manchester.ac.uk}
 \affiliation{Department of Computer Science, The University of Manchester, Manchester, U.K.}
\author{Xiao-Jun Zeng}
\affiliation{Department of Computer Science, The University of Manchester, Manchester, U.K.}
\author{Huawen Liang}
\affiliation{Department of Physics, University of Science and Technology of China, Hefei, China}
\author{Liukai Yu}
\author{Junjun Zheng}%
 \email{99zhengjunjun@163.com}
 \affiliation{School of Economics and Management, Wuhan University, Wuhan, China}

\date{Accepted on 15 July 2024 in \textit{Physical Review E}}

\begin{abstract}
This research investigates the impact of dynamic, time-varying interactions on cooperative behaviour in social dilemmas. Traditional research has focused on deterministic rules governing pairwise interactions, yet the impact of interaction frequency and synchronization in groups on cooperation remains underexplored. Addressing this gap, our work introduces two temporal interaction mechanisms to model the stochastic or periodic participation of individuals in public goods games, acknowledging real-life variances due to exogenous temporal factors and geographical time differences. We consider that the interaction state significantly influences both game payoff calculations and the strategy updating process, offering new insights into the emergence and sustainability of cooperation. Our results indicate that maximum game participation frequency is suboptimal under a stochastic interaction mechanism.  Instead, an intermediate activation probability maximizes cooperation, suggesting a vital balance between interaction frequency and inactivity security. Furthermore, local synchronization of interactions within specific areas is shown to be beneficial, as time differences hinder the spread of cross-structures but promote the formation of dense cooperative clusters with smoother boundaries. We also note that stronger clustering in networks, larger group sizes and lower noise increase cooperation. This research contributes to understanding the role of node-based temporality and probabilistic interactions in social dilemmas, offering insights into fostering cooperation.
\end{abstract}

\maketitle


\section{Introduction}
\label{sec:introduction}

Cooperation is a fundamental pro-social behaviour essential for the successful evolution of both biological ecosystems and human societies~\cite{griffin2003kin,fehr2003nature,nowak2011supercooperators}. The imperative to understand its origins and persistence has become more pressing in the context of global challenges such as climate change~\cite{hoegh2019human} and pandemics~\cite{bavel2020using}. Notwithstanding, cooperation typically requires individual sacrifices for collective gain. It presents a dilemma where selfish individuals may be tempted to defect and exploit the efforts of others, resulting in the tragedy of the commons scenario~\cite{hardin1968tragedy}. To illustrate the contrast between individual and collective optimal behaviours, the Public Goods Game (PGG) serves as a fundamental paradigm in cooperation research. 
Particularly in comparison with the Prisoner's Dilemma (PD) games where interactions occur in pairwise patterns~\cite{rapoport1965prisoner, hamburger1973n, carroll1988iterated}, the PGG involves multi-person interactions within a group and can better capture the contributions to collective welfare.

In recent decades, the advancement of evolutionary game theory (EGT) has been bolstered by developments in network science, statistical physics, and other interconnected fields~\cite{smith1982evolution,nowak1992evolutionary,perc2017statistical}. These contributions have furnished a robust mathematical framework essential for examining the emergence of cooperation in social dilemmas.
Moreover, empirical research has shown that cooperative behaviour is one form of a more general class of moral behaviour~\cite{capraro2018grand}, and this unified framework follows personal norms that beyond the monetary payoffs~\cite{capraro2021mathematical}. In order to understand which personal norms are more likely to be internalized, scholars have started applying EGT also to study the evolution of other moral behaviours, such as honesty in the sender–receiver game~\cite{capraro2020lying} and trustworthiness in the trust game ~\cite{kumar2020evolution}. Considering that real interactions are usually structured and affect behavioural decision-making, evolutionary games on graphs utilize nodes to represent individuals and links to denote interactions, allowing us to observe the evolutionary dynamics of cooperation and other moral behaviours in structured populations ~\cite{szabo2007evolutionary, capraro2018grand}.
With this approach, it has demonstrated that certain types of population structures, such as lattice~\cite{szolnoki2009topology}, small-world~\cite{abramson2001social}, scale-free~\cite{santos2008social}, are conducive to fostering cooperation. This phenomenon is identified as network reciprocity~\cite{nowak2006five}, a concept highlighting that individual interaction topologies are influenced by physical or social connections rather than occurring in a well-mixed population setting~\cite{nowak2010evolutionary,perc2013evolutionary,allen2017evolutionary}. 
Apart from the aforementioned network structure, heuristic strategies have also been instrumental in modifying individual interactions, such as migration~\cite{chen2012risk,ren2021evolutionary,fahimipour2022sharp}, volunteering~\cite{hauert2002volunteering,szabo2002phase,semmann2003volunteering}, and reputation-based partner choice~\cite{fu2008reputation,ren2023reputation}. For example, the volunteer mechanism within the spatial PGG shows `loners' opting out, earning a steady and modest income. It is important to note that while their non-participation and external benefits affect game payoffs, these strategies do not impact the dynamics of learning individual strategies.

In much of the research concerning evolutionary games on networks, a common simplification posits that individuals engage in uninterrupted interaction, participating in repeated games. This implies that deterministic interactions among neighbouring individuals are persistently active. However, real-world situations frequently diverge from this assumption. In practice, not all potential interaction relationships remain continuously operational. Instead, these interactions can be intermittent, displaying characteristics that fluctuate over time~\cite{ghosh2022synchronized}. This observation highlights the discrepancy between the idealized models commonly used in theoretical research and the more complex, dynamic nature of real-world interactions. To bridge this gap, research has begun to incorporate the time dimension to examine how the dynamic nature of time-varying interactions affects the tendency of coupled systems to achieve coherency~\cite{holme2019temporal,holme2012temporal}. 

One typical framework corresponds to the cases where the structural evolution of the graph alters the interaction in a way that switches it on and off over the course of time, such as the coevolutionary games~\cite{pacheco2006coevolution,perc2010coevolutionary} or temporal networks~\cite{holme2012temporal,hiraoka2020modeling}. In this case, the occurrence of individual interactions is fully dependent on the existence of links and acts on the strategy updating process. 
Specifically, coevolutionary games represent a coupled process in which interaction structures are fluid and evolve in tandem with individual strategies ~\cite{ebel2002coevolutionary}. As predicted by most coevolutionary game theory models~\cite{tanimoto2013difference, zheng2022evolution}, laboratory experiments based on dynamic networks provide clear evidence that the rapid making and breaking of social ties promotes cooperation and leads to greater degree heterogeneity ~\cite{rand2011dynamic, jordan2013contagion}. However, the frequency of interaction changes in these studies is mainly strategy- or payoff-driven, with less consideration of temporal effects. The temporal network (also known as a time-varying network) is a type of network representation that considers the time dimension of dynamic interactions, and discusses the times when edges are active as an explicit element~\cite{holme2012temporal}. 
As a popular family of generative temporal networks, activity-driven (AD) modeling~\cite{perra2012activity, starnini2013topological} describes the instantaneous and fluctuating dynamics through a time-invariant function called activity potential, which encodes the probability per unit time that nodes are involved in social interactions. 
The key element in the AD network is the formation of social interactions driven by the activity of individuals, urging them to interact with connected peers, and by the empirical fact that different individuals show varying activity patterns in terms of frequency and synchronization~\cite{ghosh2022synchronized}. Recently, some research has tried to associate activity-driven temporal networks and the evolutionary dynamics of cooperation~\cite{li2020evolution, han2019evolutionary}. 
According to activity potential, the nodes become active and generate pairwise links, which are deleted and rebuilt in the next time window. 
However, most social ties in real communities do not change frequently, while the interactions that happen on links are impermanent with temporal characters. Therefore, it is necessary to address the insight of individual activity into the construction of group interaction rather than the pairwise connection topology.

The second category includes systems where the time-dependency of interactions is driven by external factors, such as environmental changes. Research in this area typically models scenarios where nodes interact randomly, even with established links among neighbours, and the probability of interaction is not uniform.  The exploration of stochastic interactions within evolving games has been a crucial component of these studies~\cite {li2016changing,li2021evolution}. Individuals stochastically participate in PDGs with direct neighbours based on specified probabilities, where interaction frequency, governed by a certain distribution, only impacts game participation and not strategy updates. Nevertheless, a notable limitation of these studies is the frequent oversight of the physical features that adhere to specific spatial-temporal rules within the dynamics of interactions. This oversight is particularly evident when considering individuals with varying circadian rhythms or those situated in different geographical locations~\cite{edgar2012peroxiredoxins}. Such disparities can result in different periods of inactivity between game rounds and difficulties in synchronizing gameplay due to factors like jet lag. 
Pioneered by Kuramoto who considered the case where each oscillator is coupled to all the others~\cite{kuramoto1975self, kuramoto1984chemical}, synchronization, as an emerging phenomenon of a large population with dynamically interacting units, has been widely discussed in coupled systems. 
In the framework of complex networks where each node is considered as a Kuramoto oscillator~\cite{rodrigues2016kuramoto}, the emergence of collective synchronization with oscillating behaviour is affected by interaction topologies that are fixed in time or time-varying~\cite{arenas2008synchronization, ghosh2022synchronized}. So far, synchronized behaviours~\cite{boccaletti2002synchronization} have been mostly studied in the limit to structural properties of static networks (e.g. the degree distribution~\cite{gomez2007paths}), while relatively few applications have been conducted on time-varying intrinsic parameters (e.g. the natural frequency~\cite{petkoski2012kuramoto}) of dynamic networks. This highlights the necessity of comprehending how the synchronization developed in different temporal interaction dynamics influences the evolution of cooperation in spatial PGG.

Building upon volunteering and stochastic interaction concepts, our study introduces temporal interaction mechanisms to explore the effects of sporadic individual interactions on group cooperation. In this extended PGG framework, individuals can choose between cooperation or defection strategies and be either active or inactive. On the basis of the individual active state, we have refined the allocation of public goods benefits, ensuring inactive individuals receive a basic allowance only in the presence of active cooperators. Additionally, we established two individual activity patterns by creating state variables for each individual. These variables follow random distributions across time and space, introducing a time lag that renders interactions periodic over time and varied across regions.

\begin{figure*}[htbp]
 \centering
   \includegraphics[width=0.95\linewidth]{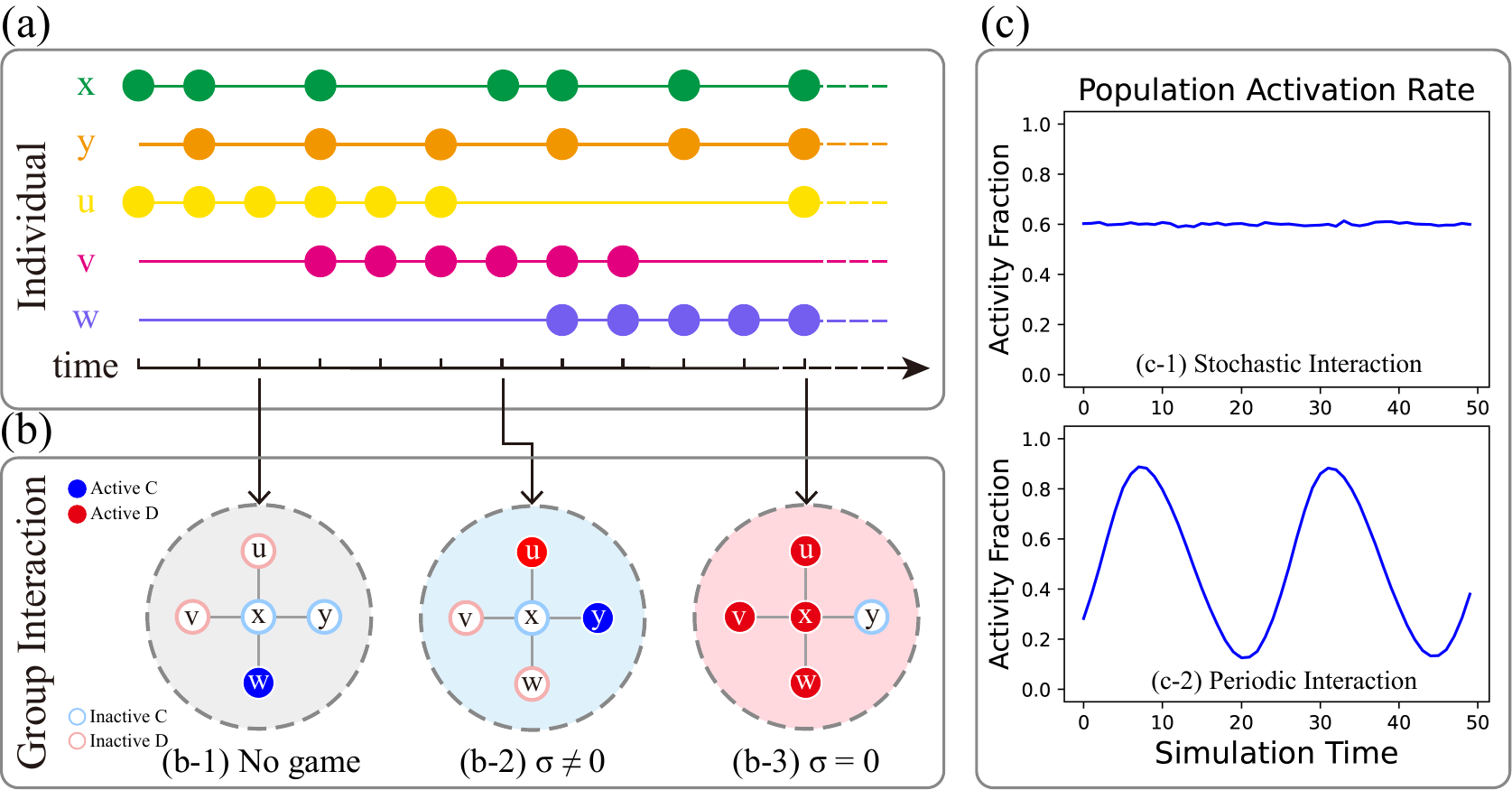}
   \caption{Schematic of temporal interaction mechanism. (a) Active patterns of five individuals, represented by solid circles of various colours. Agents $x$ and $y$ engage in stochastic interactions, while agents $u$, $v$, and $w$ follow periodic interactions. (b) Illustrates game payoff scenarios: active cooperators and defectors as filled blue and red circles, inactive ones as light blue and red empty circles. In (b-1), only $w$ is active, leading to no game and zero gain. (b-2) represents the typical situation where the active cooperator $y$ contributes to public goods, and inactive ones receive a basic allowance.} (b-3) shows zero payoff due to no active cooperators. (c) Details the calculation of the population activation rate.
   \label{fig:activity patterns}
\end{figure*}

This work offers a two-fold contribution. Firstly, we find that high-frequency game interaction is not essential for the prevalence of cooperation; instead, there is an optimal interaction frequency range that maximizes cooperation levels. Cooperation necessitates individuals being synchronously active within a localized area. Time differences, while hindering the spread of cooperation, promote the formation of highly cooperative clusters. 
Secondly, distinguishing from studies with pairwise interactions on temporal networks, we present a deeper understanding of the spatial dynamics affecting the emergence of clusters in self-organizing populations through typical snapshots of temporal group interactions and the comparison of heterogeneous network structure.
These insights could have significant implications for fostering cooperation in social and information networks.

This paper is structured as follows: Section~\ref{sec:models} outlines the models under stochastic and periodic interaction mechanisms. Section~\ref{sec:results} details numerical simulation results. Conclusions and discussions are drawn in Section~\ref{sec:discussion}. Finally, we formulate an extended pair-approximation model and provide theoretical analyses in the Appendix.

\section{Models}
\label{sec:models}

\subsection{Public Goods Game}
We consider the spatial PPGs on a $L^{2}$ square lattice with periodic boundary conditions. Each individual is situated at a node and interacts within the von Neumann neighbourhood, where $G=k+1$ represents the group size, and $k=4$ indicates the node degree. Participants can adopt one of two strategies: cooperation (C), where cooperators contribute a quantity $c$ to the common pool or defection (D), where defectors contribute nothing, opting to free-ride on the public goods. For simplicity, we standardize the investment cost $c$ to one.

\subsection{Temporal Interaction}
Considering scenarios where individuals frequently miss games due to various time-related factors, we introduce an exogenous time-dependent state variable $a_x(t) \in \{0,1\}$ to represent the interaction state of individual $x$ at time $t$. Active individuals ($a_x(t)=1$)  choose to contribute to public goods and revise strategies based on their payoff differential with neighbours. In contrast, inactive individuals ($a_x(t)=0$) neither participate in games nor change strategies, but they receive a small, fixed portion $\sigma$ of the public resources. This is due to the non-excludability and positive externalities of public goods, preventing the exclusion of non-contributors from accessing them. In each PGG group, total benefits are generated exclusively from active cooperators' contributions. After deducting the shares for inactive individuals, the residual amount is equally divided among all active members. Notably, different from the loners' scenario~\cite{szabo2002phase}, inactive individuals receive no benefits in the absence of active cooperators.

To investigate the impact of activity patterns on evolutionary dynamics, this study considers two representative distributions for activation interaction rules, namely stochastic interaction and periodic interaction. These rules introduce a degree of randomness and heterogeneity into the evolutionary timeline. Fig.\ref{fig:activity patterns} illustrates these individual activity patterns. The population activation rate is defined as the ratio of the number of activated individuals at time $t$ to the total population size. These two different interaction rules can be captured as below.

Stochastic interaction: As depicted in Fig.\ref{fig:activity patterns}(a), stochastic interaction (individuals $x$ and $y$) characterizes scenarios where human behaviour is random at any given time. We assign a probability $p$ for each individual to become active in group interactions, where $p \in [0,1]$. The number of activations $A_b$ in which an individual participates over the total simulation time $T$ follows a binomial distribution $A_b \sim B(T, p)$. 
The probability of $A_b$ taking value $n$ is $P(A_b=n)=\binom{T}{n} p^n (1-p)^{T-n}$ for $n=0,1,2\dots$, and the scenario of $p=1$ represents full interactions, reverting our model to the baseline spatial public goods game~\cite{szolnoki2009topology}.

Periodic interaction: Accounting for human circadian rhythms and location, periodic interaction alternates between engagement and rest periods as depicted in Fig.\ref{fig:activity patterns}(a) and individuals $u$,$v$,$w$. For each individual, the number of activations $A_p$ within the time $[t, t+\tau)$  follow a Poisson distribution with parameter $\lambda$ , as $A_p \sim P(\lambda)$. Here, $\tau$ is a unit like a day or week, and $\lambda$ represents the average number of active participations per time interval. The probability of $A_p$ taking value $n$ is $ P(A_p = n) = \frac{\lambda^n e^{-\lambda}}{n!} $ for $n=0,1,2\dots$. Thus, the average activation probability $\bar{p} = \lambda / \tau$, simplified to $p$ for further discussion. Additionally, a random variable $\epsilon \sim N(\mu, \sigma^2)$ is introduced to indicate the time lags effect on interaction states across different areas. A node close to the network centroid is arbitrarily selected as a reference point, with its time serving as the basis for simulation time. The mean value $\mu$ of $\epsilon$ corresponds to the Euclidean distance $d$  between an individual and this reference node:
\begin{equation}
    d(i,j) = \sqrt{(x_i-x_j)^2 + (y_i-y_j)^2}
    \label{eq:d[ij]}
\end{equation}
where $2\mu d_{max}=\tau$ ensures that the maximum time difference between the farthest individuals does not surpass the time interval. The local time for each agent is set to $t-\epsilon$. All probability is assumed to be independently and identically distributed among different individuals.

\subsection{Evolutionary dynamics}
Simulations are conducted using Monte Carlo methods. Initially, each individual on site is randomly assigned a strategy with equal probability, and all actively participate in games. Subsequently, an activation schedule is generated for each individual according to the temporal interaction rule, defining their state at any point during the evolutionary process. For periodic interaction, we set $\tau=24$, $\mu=12\sqrt{2}L$ and $\sigma=2$ to simulate daily human activities. The PGG then advances through a sequence of elementary steps.

Step 1 (Game Interaction): A random individual $x$ is selected as the focal player to participate in the PGGs occurring at its site and neighbouring sites. Within each group $g$, let $i_c$ ($i_d$) denote the number of active cooperators and defectors, respectively, and $i=i_c+i_d$ represents the total number of active neighbours. If $a_x+i \leq 1$, the game does not take place, and all individuals receive a payoff of 0 (Fig.\ref{fig:activity patterns}(b-1)). 
In other scenarios, the multiplication factor $r \in (1, G)$ determines the total contributions from cooperators. With the experimental evidence in~\cite{capraro2014heuristics,hauert2003prisoner}, an increase in $r$ leads to a diminished dilemma strength and favours cooperation. Accordingly, the payoff of active cooperator (AC) and active defector (AD) in one game are given as
\begin{equation}
    \pi_{AC} = \frac{r(i_c+1)-\sigma(k-i)}{i+1}-1,
    \label{eq:pai_ac}
\end{equation}
\begin{equation}
    \pi_{AD} = \frac{ri_c-\sigma(k-i)}{i+1}.
    \label{eq:pai_ad}
\end{equation}

Both inactive cooperator (IC) and inactive defector (ID) receive a payoff
\begin{equation}
    \pi_I = \begin{cases}
        \sigma, i_c \geq 1 \\
        0, i_c=0
    \end{cases},
    \label{eq:pai_i}
\end{equation}
where  $\sigma \in (0,r-1)$ represents the inactivity payoff. Specifically, when there are no active individuals investing in public goods, all AD, IC and ID have no benefits, as illustrated in Fig.\ref{fig:activity patterns}(b-3). The overall payoff at time step $t$  is the sum of all payoffs from the games participated in, calculated as $\Pi_x=\sum_{g=1}^{G}\pi_{x}^{g}$.

Step 2 (Update of Strategy and State): Following game participation, individuals update their strategies asynchronously in a randomly sequential manner. Initially, individual $x$ randomly selects one of its neighbours $y$, who acquires its payoff $\Pi_y$. Then, $x$ adopts $y$'s strategy with a probability determined by the Fermi function~\cite{szabo2007evolutionary}:
\begin{equation}
    f(s_x \to s_y)=\frac{a_x(t)}{1+e^{[-(\Pi_y-\Pi_x)/K]}},
    \label{eq: fermi}
\end{equation}
where $K$ measures noise intensity. With $K \to 0$, individuals with higher payoffs easily dominate strategy-wise over those with lower payoffs. Conversely, for $K \to \infty$, strategy imitation becomes indiscriminate of payoffs, indicative of weak selection~\cite{ohtsuki2006simple}. We set $K=0.1$, favouring the imitation of higher payoff strategies, though noise allows for the occasional adoption of less successful ones. It is important to note that strategy updates are exclusive to \textit{active} individuals, while \textit{inactive} ones keep their strategies but may still influence others (Eq.\ref{eq: fermi}). Each individual $x$ updates its state to $a_x(t+1)$ for the next game round. The process concludes with $L^{2}$ iterations, allowing every individual the opportunity to revise their strategy and state once on average, completing a full Monte Carlo step(MCS).

\section{Results}
\label{sec:results}
In the following subsections, we present simulation results conducted on a square lattice of size $10^4$. The key value $\rho_c$, characterizing the frequency of cooperators in the population, is derived by averaging over the last $10^3$ generations, following more than $10^4$ time steps. To ensure robustness and minimize variability, the final steady states are obtained by conducting up to 10 independent realizations.

\subsection{Overview of temporal interaction mechanism}

\begin{figure}
   \centering
    \includegraphics[width=\columnwidth,keepaspectratio]{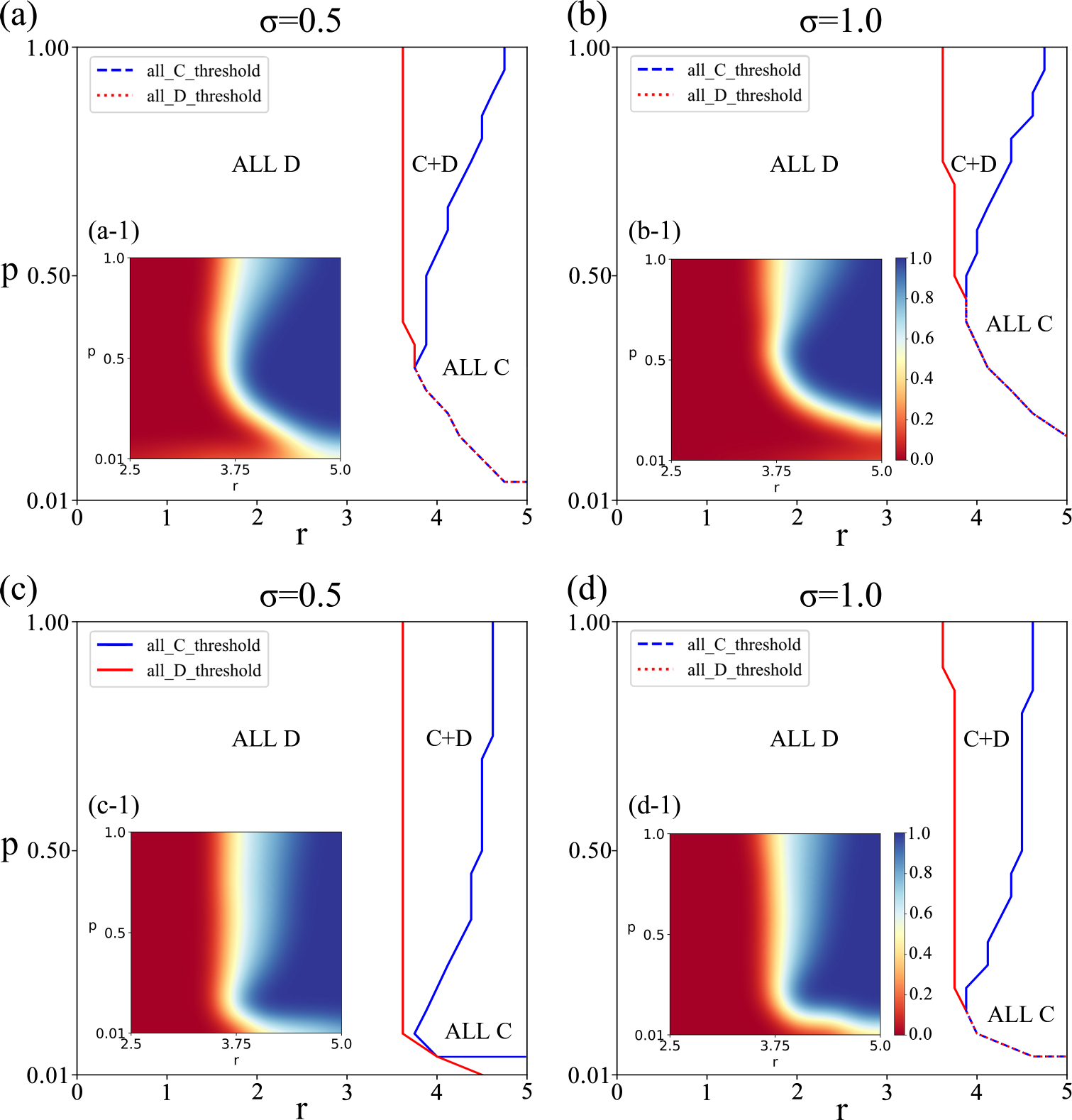}
   \caption{$r-p$ phase diagrams of the spatial public goods game as obtained for different inactive income $\sigma=0.5$ (left column) and $\sigma=1$ (right column). Subfigures (a) and (b) represent stochastic interaction, (c) and (d) represent periodic interaction. The red (blue) line represents the phase transitions between the mixed C+D and homogeneous D (C) states, and the solid (dashed) lines signify the continuous (discontinuous) phase transitions. The inserted colourmap $(^*-1)$ of each panel shows the equilibrium fraction of cooperators with $r$ and $p$. From red to blue, the colour bar indicates that the cooperation level changes from 0 to 1 accordingly. All results are obtained for $K=0.1$ and $t=10^5$.}
   \label{fig:2}
\end{figure}

We begin by providing a comprehensive analysis of the transition phenomenon, focusing on the multiplication factor $r$ and activation probability $p$. Fig.\ref{fig:2} presents $r-p$ phase diagrams illustrating the cooperation level $\rho_c$ under both stochastic and periodic interaction mechanisms. Notably, although a portion of the investment returns in public goods is distributed to inactive individuals, the threshold for cooperators to surpass defectors is significantly lower than the null model. This implies that temporal interaction mechanisms are conducive to the flourishing of cooperation. Moreover, at $\sigma=1$, the proportion of defectors is reduced compared to the volunteering model~\cite{szabo2002phase}, increasing the likelihood of the group reaching a global cooperation steady state. However, the fixed payoff $\sigma$ of inactive individuals undermines cooperation. As $\sigma$ increases from $0.5$ to $1$, the transition lines between ALL D, C+D and ALL C shrink, along with a reduction in the area representing the C+D phase. Interestingly, the transition between the all-cooperator and all-defector states is discontinuous, with low activation probabilities. Conversely, under conditions of frequent interactions, the system evolves into a mixed C+D state through a continuous phase transition.

Further, there exists a non-linear interplay between multiplication factor $r$ and activation probability $p$ in fostering cooperation, with varying impacts depending on the activity pattern. As depicted in Fig.\ref{fig:2}(a) and (b), within the stochastic interaction mechanism, a negative correlation between $r$ and $p$ is observed at $p<0.5$.  Specifically, as $p$ decreases, the threshold $r$ required to shift the system from an all-defector (ALL D) to an all-cooperator (ALL C) state incrementally rises. It's important to note that cooperators fail to sustain themselves at $p<0.1$, underscoring the critical role of adequate interaction for the emergence of cooperation. In scenarios of higher activation frequency ($p \geq 0.5$), the transition from an ALL D to a mixed cooperator-defector (C+D) phase shows limited sensitivity to changes in $p$. However, the multiplier $r$ needed to achieve an ALL C phase escalates with increasing $p$, indicating a positive correlation. This suggests that, in stochastic interactions, an optimal activation probability $p$ exists at which the system can evolve into an ALL C equilibrium with the smallest required $r$.

Furthermore, periodic interactions appear to have a slightly more pronounced positive influence on cooperative behaviour. Fig.\ref{fig:2}(c) and (d) show that the C+D and ALL C regions are more extensive compared to those in stochastic interactions. Interestingly, the threshold for the emergence of cooperation is not influenced by $p$, with the ALL D threshold situated around $r=3.62$ in Fig.\ref{fig:2}(c) and $r=3.75$ in (d). However, as $p$ increases from 0.01 to 1, the required $r$ to achieve an ALL C state first decreases and then rises continuously. Even with $p<0.1$, effective cooperation can still emerge and predominate with suitable $r$ values, as seen in Fig.\ref{fig:2}(c). This indicates that small-scale activation may more effectively lead to global cooperation, a phenomenon further examined through spatio-temporal factors in Fig.\ref{fig:6}.

\subsection{Effects of stochastic interactions}

\begin{figure}
   \centering
    \includegraphics[width=\columnwidth,keepaspectratio]{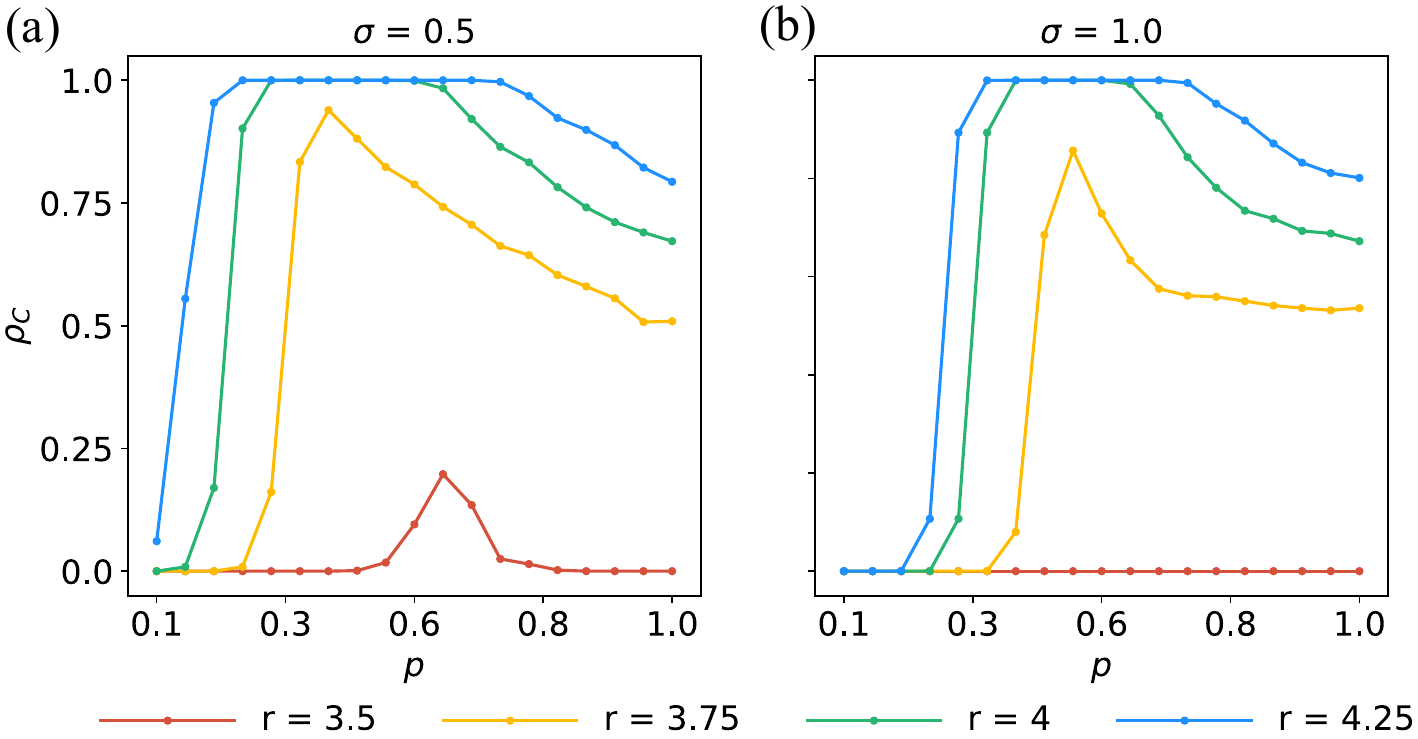}
   \caption{The cooperation level $\rho_c$ as a function of activation probability $p$ for various  multiplication factors $r$ under a stochastic interaction mechanism. Results are shown for (a) $\sigma=0.5$ and (b) $\sigma=1$. In each subplot, four distinct curves correspond to $r$ values of 3.5, 3.75, 4, and 4.25, respectively.}
   \label{fig:3}
\end{figure}

\begin{figure*}[htbp]
   \centering
    \includegraphics[width=\linewidth,keepaspectratio]{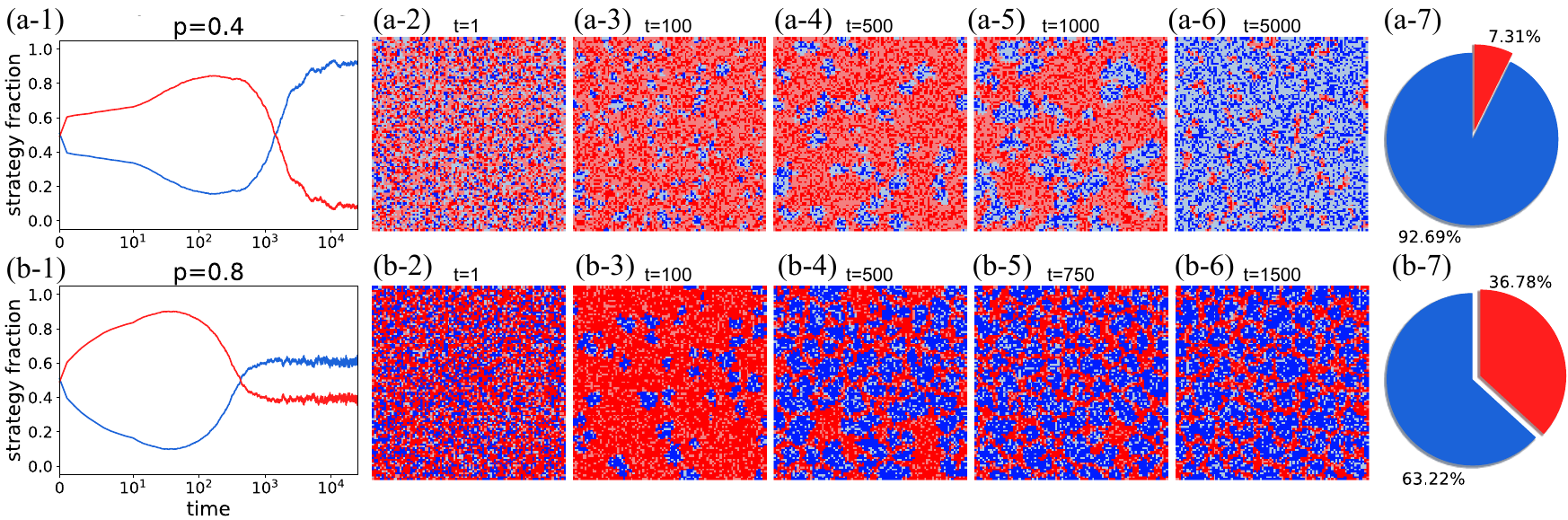}
   \caption{Evolutionary dynamics and spatial distributions of strategies under stochastic interaction mechanism. Column (*-1) illustrate the evolution process of strategy proportions over time, where the blue (red) line indicates the fraction of cooperation (defection) that contains active and inactive states, and (*-7) show the strategy fraction at the equilibrium state.} Columns (*-2) to (*-6) display snapshots at specific iterations, with $p=0.4$ and $t=1,100,500,1000,5000$ in panel (a-*), as well as $p=0.8$ and $t=1,100,500,750,1500$ in panel (b-*). Dark and light red represent active and inactive cooperators, while dark and light blue denote active and inactive defectors, respectively. All results are obtained for $r=3.75$, $\sigma=0.5$ over $t=25000$ iterations.
   \label{fig:4}
\end{figure*}

To more precisely quantify the role of stochastic interaction in promoting cooperation and determine the critical exponent, our study investigates the dependence of cooperation level ($\rho_c$) on activation probability ($p$) for varying multiplication factors ($r$), as illustrated in Fig.\ref{fig:3}.  We observe that for a fixed $r$ the proportion of cooperators initially increases with $p$, reaching an optimum, and then decreases as $p$ further escalates to 1. This pattern suggests an optimal $p$-range conducive to achieving full cooperation, indicating that maximal cooperative spread under stochastic interaction occurs at an intermediate temporality. This 'Goldilocks effect' of temporality, eqnarraying with findings in spatial PDG~\cite{li2020evolution,chen2008interaction}, demonstrates that neither too frequent nor too rare interactions are ideal for promoting cooperation. Moreover, as $p$ surpasses this optimal range, $\rho_c$ starts to decline monotonically, eventually stabilizing at a mixed cooperator-defector (C+D) strategy equilibrium (for $r = 4$ and $4.25$). Additionally, at a fixed $p$ and $\sigma$, a higher $r$ correlates with an increased $\rho_c$ at evolutionary equilibrium, where the plateau's length expands with $r$ but contracts with $\sigma$. Notably, in Fig.~\ref{fig:3}(b) with $\sigma=1$, the optimal $p$-value for fostering cooperation is significantly larger compared to $\sigma=0.5$ in Fig.~\ref{fig:3}(a). This reflects the real-world scenario where higher basic security from public goods necessitates more frequent interactions to achieve cooperative consensus. In such cases of less active, the disparity in individual gains is minimal, and the strategy learning probability calculated by Eq.\ref{eq: fermi} in each MCS is low.

In order to intuitively understand the impact of stochastic interaction on cooperation at the micro level, we examine the yellow curve in Fig.~\ref{fig:3}(a) for the activation probability of \(p=0.4\) and \(p=0.8\). Some typical snapshots are provided in Fig.~\ref{fig:4}. The overall trend shows an initial decline followed by an increase in the prevalence of cooperation in Fig.~\ref{fig:4}(*-1). Following the framework in Wang et al. (2013)~\cite{wang2013insight}, this process encompasses two phases: the enduring (END) phase and the expanding (EXP) phase. The END phase is characterized by a rapid decrease in cooperation and the emergence of large defector clusters. Cooperators, however, persist in the spaces between these clusters, protected by inactive individuals who inhibit the spread of defection, observed in Fig.~\ref{fig:4}(*-3). Next, in the EXP phase, small cooperative groups encircled by defectors start to grow. Active individuals at the irregular boundary are more likely to become cooperators due to higher benefits within cooperative clusters, leading to a rapid increase and subsequent stabilization of cooperation levels around a mixed C+D equilibrium. Interestingly, cooperation level at equilibrium is higher at smaller active probability $p$, and the spatial distribution of the strategies shows a great deal of variation in Fig.~\ref{fig:4}(*-6). As shown in (a-4) to (a-6), although the presence of inactive individuals creates a significant hurdle to forming ideal cooperative clusters, it protects the expansion of cooperative cluster, while a small number of defectors exist in unevenly distributed clusters. By contrast, high-frequency interactions allows defectors to connect and form a mesh structure. This net divides the cooperative group into small size and decreases the level of cooperators, see (b-4) to (b-6).

\subsection{Dynamics of periodic interactions based on time-lag}

\begin{figure}
   \centering
\includegraphics[width=\columnwidth,keepaspectratio]{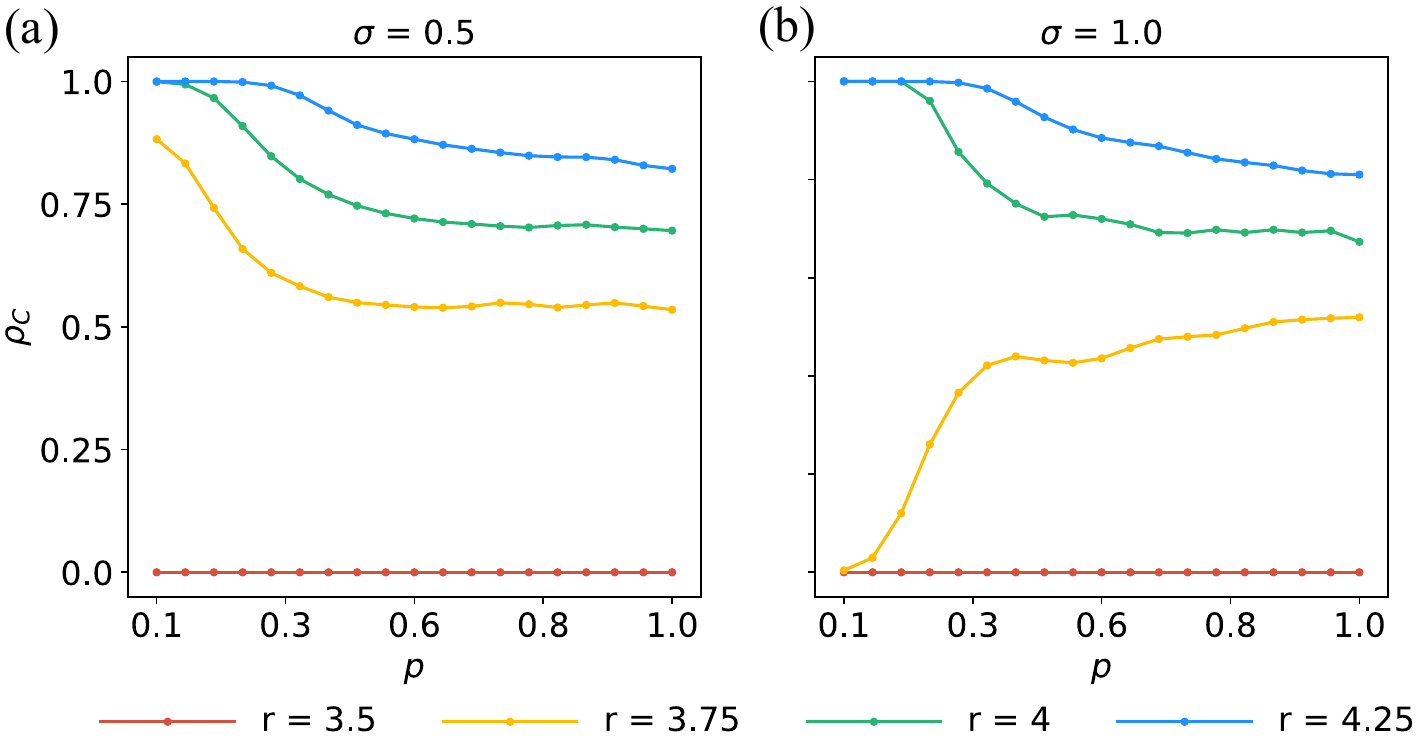}
   \caption{The cooperation level $\rho_c$ as a function of activation probability $p$ for various multiplication factors $r$ under a periodic interaction mechanism. The results are presented for (a) $\sigma=0.5$ and (b) $\sigma=1$. In each subplot, the four curves correspond to $r=3.5,3.75,4$ and $4.25$, respectively.}
   \label{fig:5}
\end{figure}

Recalling Fig.~\ref{fig:2}(c) and (d), the periodic interaction mechanism enables cooperation to prevail with a small average active probability. Fig.~\ref{fig:5} plots how $\rho_c$ varies as a function of $p$ under different values of $r$, revealing that cooperation is maximized at the combination of intermediate $r$ and low $p$. When $r$ exceeds the ALL D threshold in Fig.~\ref{fig:2}, the proportion of cooperation decreases from 1 with increasing $p$, reaching a C+D mixed equilibrium. Counterintuitively, Fig.~\ref{fig:5} shows that the inactivity payoff $\sigma$ affects the evolution of cooperation but not the final cooperation level. As $r$ increases, the downward trend of $\rho_c$ slows and stabilizes at a higher level than the equilibrium fraction in Fig.~\ref{fig:3}. An exception is the yellow line in Fig.~\ref{fig:5}(b), which presents a positive correlation between $\rho_c$ and $p$, as $r=3.75$ is the critical value for cooperation evolution under $\sigma=1$.

\begin{figure}
   \centering
    \includegraphics[width=\columnwidth,keepaspectratio]{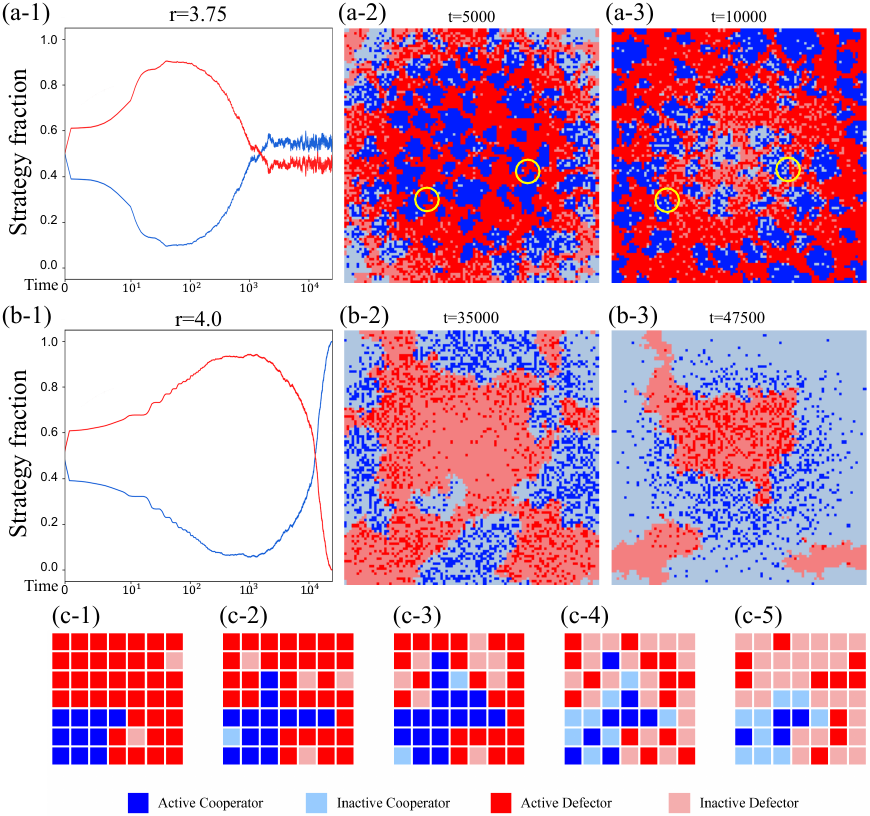}
   \caption{Evolutionary dynamics and spatial distributions of strategies under periodic interaction mechanism. Panel (a-*) are obtained for $r=3.75$, $p=0.6$ over $t=50000$ iterations, as well as $r=4$, $p=0.1$ and $t=100000$ in panel (b-*). (a-1) and (b-1) illustrate the evolution process of strategy proportions. (*-2) and (*-3) display snapshots at specific iterations $t=5000,10000$ in (a-*) and $t=35000,47500$ in (b-*).  Taking the yellow circles in (a-2) as an example, panel (c-*) visualize the strategy changes of individuals on the boundary of cluster. Dark and light red represent active and inactive cooperators, while dark and light blue denote active and inactive defectors, respectively. All results are obtained for $\sigma=1$.}
   \label{fig:6}
\end{figure}

To further explore the reason why periodic interaction shows a significant positive effect when activation frequency is small, Fig.~\ref{fig:6} records snapshots of the spatial distribution at different simulation times. As shown, the time difference leads to the phenomenon of periodic shift of the active center at the macro level, and the activity patterns are similar among individuals located in the same region. When one region is active, nearly all individuals within that region participate in games, while others far away are inactive. Comparing Fig.~\ref{fig:6}(a-*) and (b-*), we observe that higher $p$ forms many small-sized C clusters with irregular boundaries, whereas lower $p$ produces fewer C clusters with smoother boundaries.

In Figs.\ref{fig:6}(a-1) and (b-1), the fraction of cooperation decreases during the END period and rises up to 0.55 and 1, respectively, in the EXP period. With $p=0.1$ in (b-*), strategy updates are less frequent, leading to a prolonged journey to stochastic equilibrium. The trend of periodic fluctuations in $\rho_c$ mirrors the population activity rate in Fig.\ref{fig:activity patterns}(c-2), increasing initially and then declining. Fig.\ref{fig:6}(c-*) illustrates this change in cooperative clusters over time at the micro level. Active individuals at cooperative cluster boundaries can expand outward due to payoff settings, forming cross-like structures as seen in Figs.\ref{fig:6}(c-1) to (c-3). However, with the shifting active center and an increase in inactive neighbors, these active cooperators face exploitation by defectors, leading to a reduction in their numbers (Fig.\ref{fig:6}(c-4)) and eventually forming larger, more compact cooperative clusters with smoother boundaries (Fig.\ref{fig:6}(c-5)). These clusters are more resilient against defection in subsequent active periods. Consequently, sparse activity under a periodic interaction mechanism is more conducive to the formation and integration of robust cooperative clusters. In Figs.\ref{fig:6}(b-2) and (b-3), these compact cooperative clusters gradually connect and encircle defector clusters, ultimately achieving an ALL C equilibrium.

\subsection{Robustness verification}

\paragraph{Heterogeneous degree distribution and stochastic interaction.}

\begin{figure}
   \centering
   \includegraphics[width=\columnwidth,keepaspectratio]{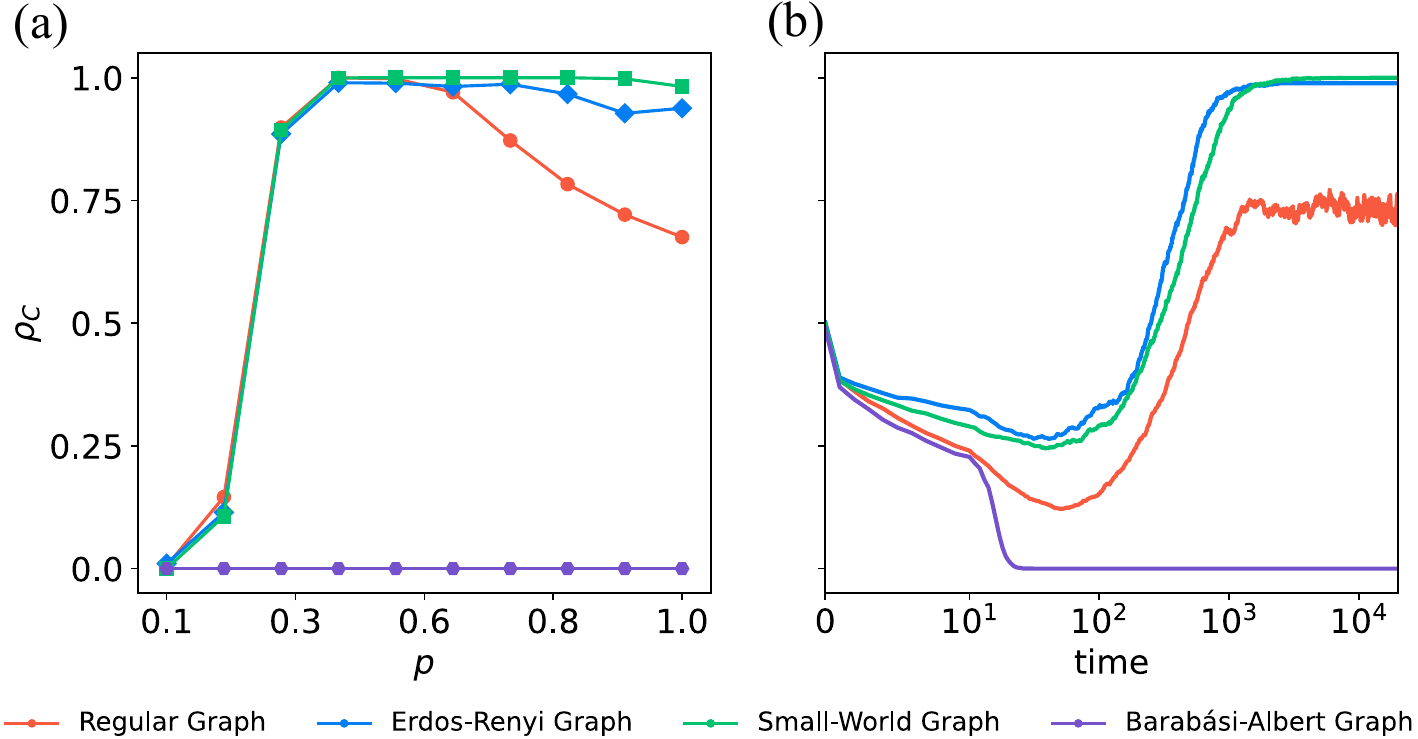}
   \caption{Effects of different network structures with the stochastic interaction. (a) The cooperation level $\rho_c$ as a function of activation probability $p$ on regular, ER, SW and BA graphs with $r=4$. (b) The evolutionary process for the fractions of cooperation strategies $\rho_c$ over time at $r=3.75$ and $p=0.6$. All results are obtained for $\sigma=0.5$ and $t=20000$.}
   \label{fig:7}
\end{figure}

Consistent with the square lattice network, we generate Erdos-Renyi random network (ER)~\cite{erdHos1960evolution}, Watts-Strogatz small-world network (SW)~\cite{watts1998collective} and Barabási-Albert scale-free network (BA)~\cite{barabasi1999emergence} with the average degree ⟨k⟩ = 4.
As shown in Fig.\ref{fig:7}, the clustering is a double-edged sword. On the one hand, the optimal platform for $p$ is not observed in the ER and SW graphs, which means high-frequency stochastic interactions not constrain the flourish of cooperation, see Fig.\ref{fig:7}(a). 
Since the ER and SW networks have small average shortest path lengths and large local clustering coefficients, the hubs can serve as common connections to mediate the short path lengths between other edges. In other words, even distant individuals belonging to different regions can connect via the hubs that work like social media platforms, thus facilitating the flow of information and capital beyond the limitations of distance, enabling individuals to make decisions with a more holistic perspective, and generating collective behaviours~\cite{bakshy2015exposure}. A recent study of temporal interactions had led to the similar conclusion that the appropriate absence of hubs boosts cooperation~\cite{meng2024promoting}, which supports our finding. By contrast, the lack of cliques and clustering coefficient equal to zero is are the primary reason that nonlinear temporality is only seen in regular networks. As shown in Fig. \ref{fig:4}(b-6), the defectors have formed a net structure to separate the cooperators into unconnected clusters with frequent interactions. The social sphere for each individual in regular network is limited by Information Cocoons~\cite{sunstein2006infotopia}, thereby bias decision outcomes and it is difficult to connect different clusters to form synergies.

Besides, cooperators have a similar time gaining footholds but smaller fluctuations to reach higher cooperation level at equilibrium in ER and SW networks, see Fig.\ref{fig:7}(b), because the social diversity~\cite{santos2008social} is more conducive to the formation of cooperative clusters. 
On the other hand, the strong heterogeneity of degree and low clustering coefficient in BA graph fully inhibits cooperation, as inactive individuals with high-degree can take free-rides extensively, destroying fragile cooperative structures early in the evolutionary process. The similar phenomenon has been found in periodic interactions.

\paragraph{Different group size and periodic interaction.}

\begin{figure}
   \centering
    \includegraphics[width=\columnwidth,keepaspectratio]{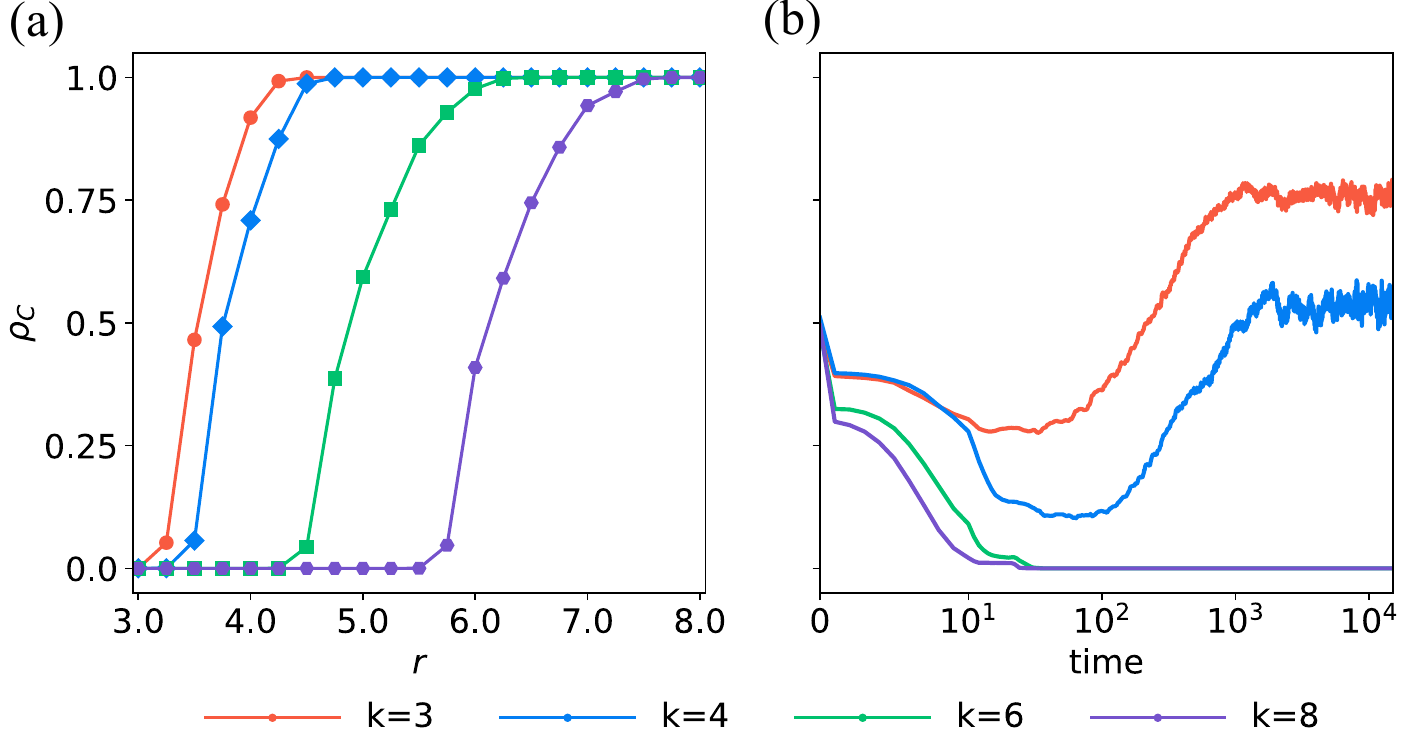}
   \caption{Effects of different group size with the periodic interaction. (a) The cooperation level $\rho_c$ as a function of multiplication factor $r$ on regular graphs with k = 3, 4, 6, 8. (b) The evolutionary process for the fractions of cooperation strategies $\rho_c$ over time at $r=3.75$ and $p=0.6$. All results are obtained for $\sigma=0.5$ and $t=15000$.}
   \label{fig:8}
\end{figure}

In agreement with the fact in Fig.\ref{fig:6} that cooperation requires a sufficient number of active individuals in a localized area, we may argue that larger groups facilitate the survival of cooperators. Results presented in Fig.\ref{fig:8} evidence clearly that group size $G=k+1$ plays a decisive role by the evolution of cooperation in PGG on regular graphs with different node degree k = 3, 4, 6 and 8. 
From a per capita perspective, the marginal return $r/G$ required for cooperation to emerge and dominance decreases slightly as the group size is enlarged (see Fig.\ref{fig:8}(a)), which supports the preceding statement. This fact has long been confirmed by several experimental studies~\cite{barcelo2015group,capraro2014heuristics,pereda2019group}, and they argued that the group size has a positive effect on cooperation in the PGG only when it generates a linear increase of the benefit $r$ or a decrease of the cost $c$ of cooperation. However, by leaving both the benefit and cost constant, the group size has a pure negative effect on cooperation in Fig.\ref{fig:8}(b). The collective gains are susceptible to be free ridden in large-size group interactions, resulting in a lower cooperation level at equilibrium, and this fact has also been confirmed by several experimental studies~\cite{barcelo2015group, isaac1984divergent}.

\begin{figure}
   \centering
    \includegraphics[width=\columnwidth,keepaspectratio]{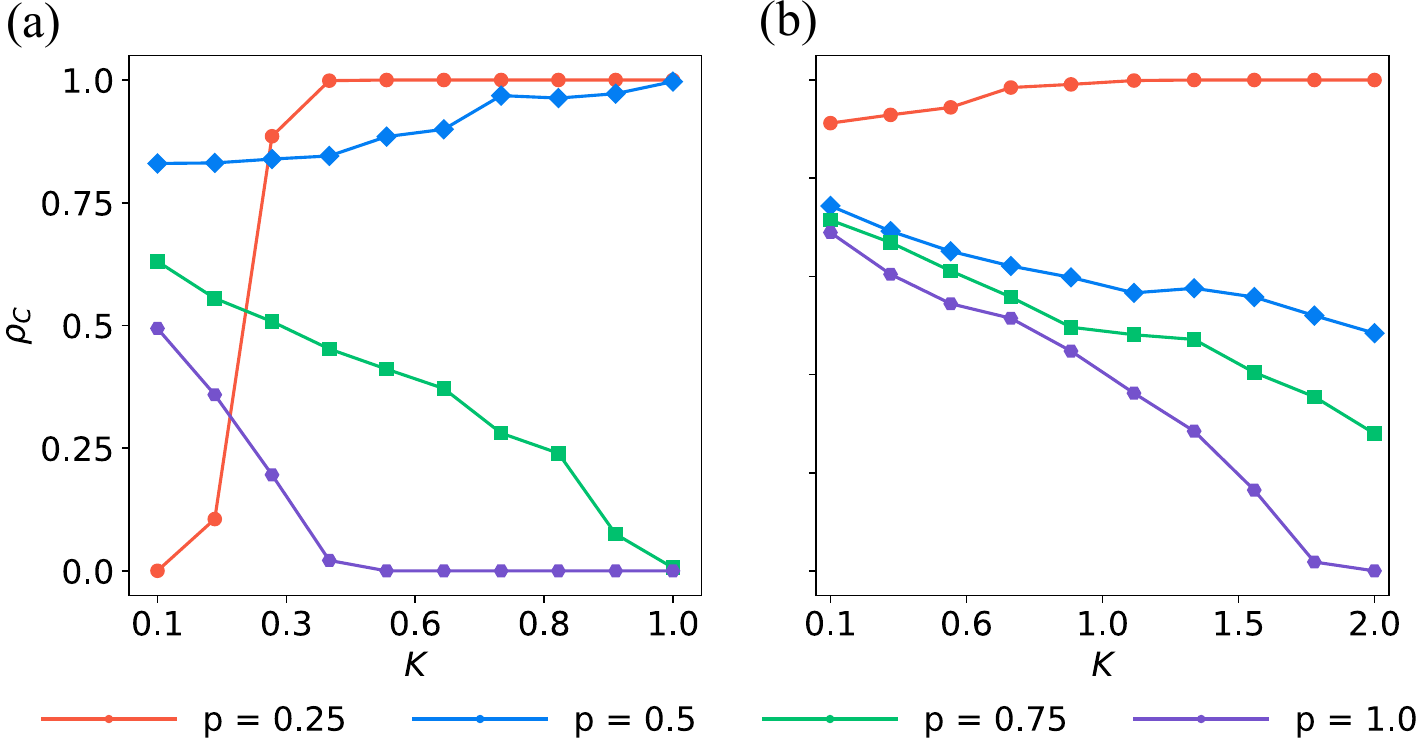}
   \caption{Effects of noise $K$ on the cooperation level $\rho_c$ under different activation probability $p$. Results are obtained for (a) stochastic interaction with $r=3.75$ and (b) periodic interaction with $r=4.0$. In each subplot, four colour curves correspond to $p$ values of $0.25$, $0.5$, $0.75$ and $1$, respectively. All results are obtained for $\sigma=0.5$ and $t=25000$.}
   \label{fig:9}
\end{figure}

\paragraph{The effect of noise on the evolution of cooperation.}
As described in Section II, the noise $K$ characterizes the irrationality during the process of strategy imitation. Therefore, it is of great significance to explore the robustness of dynamic systems over different parameter $K$, especially in the case of weak selection(high noise). 
The robustness verification of stochastic interaction in Fig.\ref{fig:9}(a) and periodic interaction in  Fig.\ref{fig:9}(b) both indicate an optimal active threshold for facilitating cooperation in different noise environments, which again supports our findings of the temporal ‘Goldilocks effect’. 
As $p$ increases to $1$, the proposed model reverts to static networks, and an increase in $K$ leads to a monotonically decreasing level of cooperation, in line with the previous work \cite{szolnoki2009topology}. By contrast, the decreased interaction frequency can instead encourage cooperation as K increases, see the curves for $p =0.25,0.5$ in Fig.\ref{fig:9}(a) and $p=0.25$ in Fig.\ref{fig:9}(b).
This implies that cooperative behaviour can thrive on inactivity when decision-making is irrational and a similar phenomenon was found in \cite{guan2007effects} concerning activity teaching. These results prove that noise and active participation have a nontrivial relationship in the evolution of cooperation.

\section{Discussion}
\label{sec:discussion}
In real society, individuals exhibit diverse active patterns, alternating between periods of activity and rest. This variability, influenced by differences in biological clocks and geographic locations, means that individuals in different states seldom interact simultaneously. Consequently, actual partnership interactions are dynamic in nature and characterized by temporal variations. In this context, we propose two temporal mechanisms for participation in spatial public goods games: stochastic interaction and periodic interaction. An individual's state not only influences the payoffs from the game but also affects the strategy updating process. Furthermore, acknowledging the non-exclusive nature of public goods, we introduce a rule for sharing collective benefits. Specifically, inactive individuals are allowed to free-ride only when there are contributions from active cooperators.

The results demonstrate that temporal interaction significantly enhances cooperation, with the level of enhancement being greatly influenced by interaction patterns and activation probability. Notably, the transition line from a mixed C+D to an ALL C state is quasi-concave, revealing an optimal combination of $p$ and $r$ that maximizes cooperation. In the realm of stochastic interactions, an intermediate $p$ value creates an environment conducive to cooperative behaviour, suggesting that a moderate frequency of interactions, rather than continuous engagement, is more effective in promoting cooperation.  Conversely, periodic interactions, characterized by synchronized activities within small regions, positively influence cooperative evolution. From a macro perspective, temporal disparities lead to shifts in the active centre, thereby limiting the spread of cooperation to inactive areas. However, at the micro level, the inactivity of neighbors curtails the outward expansion of cooperative structures, thereby facilitating the formation of more robust cooperator clusters, particularly at lower $p$ values. Furthermore, the combination of spatial and temporal heterogeneity is discussed. It reveals that the strong heterogeneity of degree distribution tends to inhibit cooperation, whereas a larger group size promotes cooperation and moderate some defectors inside.

In summary, our study explores the impact of interaction frequency and synchronization on the evolution of cooperation in spatial PGG. Utilizing Monte Carlo simulations and theoretical calculations, we have demonstrated that maintaining cooperation requires an appropriate balance of interaction frequency and local synchrony. This work potentially positions temporal interaction as an alternative mechanism for enhancing cooperation. It also provides great potential for wider applications to investigate the evolution of truth-telling, trust, or other moral behaviours from a temporal network perspective. Future work could explore individual interaction preferences through a utility function, investigate the boundary conditions of the moral phenotype particularly focusing on temporal interactions, or guide the orchestration and optimization interaction sequences in practical scenarios, which will continue to be a focal topic in modern science.

\begin{acknowledgments}
This research was supported by the National Natural Science Foundation of China (Grant Nos. 72171180).
\end{acknowledgments}

\appendix 
\renewcommand{\theequation}{A\arabic{equation}} 
\setcounter{equation}{0} 
\renewcommand{\thefigure}{A\arabic{figure}} 
\setcounter{figure}{0}

\section{Extended pair-approximation approach}
Based on the pair approximation approach \cite{ohtsuki2006simple, sun2023state}, let $\rho_x$ and $\rho_{xy}$ denote the fraction of strategy X and XY pairs in the population, where $x,y \in {C, D}$. Let $q_{x|y}$ represent the conditional probability to find a X-player given that the adjacent node is occupied by a Y-player. Based on these definitions, the following relations
\begin{eqnarray}
    \rho_x+\rho_y &=& 1 \nonumber \\
    q_{x|y}+q_{y|y} &=& 1 \nonumber \\
    \rho_{xy} &=& \rho_y q_{x|y} \nonumber \\
    \rho_{xy} &=& \rho_{yx} \nonumber
\end{eqnarray}
imply that the whole system can be described by only two variables $\rho_x$ and $q_{x|y}$ in the pair approximation approach. Specifically, $\rho_d=1-\rho_c$, $q_{d|c}=1-q_{c|c}$, $q_{c|d}=\frac{\rho_{cd}}{\rho_d}=\frac{\rho_c(1-q_{c|c})}{1-\rho_c}$, $q_{d|d}=\frac{1-2\rho_c+\rho_c q_{c|c}}{1-\rho_c}$, $\rho_{dc}=\rho_{cd}=\rho_c q_{d|c}=\rho_c(1-q_{c|c})$ and $\rho_{dd}=\rho_d q_{d|d}=1-2\rho_c+\rho_c q_{c|c}$.

We first consider the accumulated payoff of AD-player who has $i_c (i_d)$ AC (AD)-neighbors, and $i=i_c+i_d$ represents the number of active individuals among the $k$ nearest neighbors. According to Section \ref{sec:models}, the total payoff of focal player is accumulated from $k+1$ involved PGGs, but here we assume that the payoff is merely determined by a single PGG. This simplification makes the pair approximation more convenient while causing minor modifications in the systems dynamics~\cite{wang2020spatial}. In addition, AD gains nothing when $i=0$ or $i_c=0$. Therefore, the expected payoff of an active defector in a PGG is
\begin{eqnarray}
    \overline{\Pi}_{AD} &=& \sum_{i=1}^{k} \binom{k}{i}p^i(1-p)^{k-i} \sum_{i_c=1}^{i} \binom{i}{i_c} q_{c|d}^{i_c}  q_{d|d}^{i-i_c} \nonumber \\
     & &\times \left(\frac{r i_c - \sigma(k-i)}{i+1}\right).
    \label{eq:A1}
\end{eqnarray}

Similarly, the expected payoff of AC, ID and IC is respectively given as
\begin{eqnarray}
    \overline{\Pi}_{AC} &=& \sum_{i=1}^{k} \binom{k}{i}p^i(1-p)^{k-i} \sum_{i_c=0}^{i} \binom{i}{i_c} q_{c|c}^{i_c}  q_{d|c}^{i-i_c} \nonumber \\
     & &\times \left(\frac{r(i_c+1) - \sigma(k-i)}{i+1}-1\right),
    \label{eq:A2}
\end{eqnarray}
\begin{equation}
    \overline{\Pi}_{ID} = \sum_{i=2}^{k} \binom{k}{i}p^i(1-p)^{k-i} \sum_{i_c=1}^{i} \binom{i}{i_c} q_{c|d}^{i_c}  q_{d|d}^{i-i_c} \sigma,
    \label{eq:A3}
\end{equation}
\begin{equation}
    \overline{\Pi}_{IC} = \sum_{i=2}^{k} \binom{k}{i}p^i(1-p)^{k-i} \sum_{i_c=1}^{i} \binom{i}{i_c} q_{c|c}^{i_c}  q_{d|c}^{i-i_c} \sigma.
    \label{eq:A4}
\end{equation}

Next, if the selected AD-player imitates a C-neighbor successfully, then $\rho_c$ increases by $1/N$  with the transition probability
\begin{eqnarray}
    T^{+}(\Delta \rho_c=1/N) &=& \rho_d q_{c|d} [p f(AD \rightarrow AC) \nonumber \\
    & &+(1-p)f(AD \rightarrow IC)],
    \label{eq:A5}
\end{eqnarray}
where $N$ is the size of population, which equal to $L^2$. And the transition probability that $\rho_c$ decreases by $1/N$ because a selected AC imitates a D-neighbor successfully is 
\begin{eqnarray}
    T^{-}(\Delta \rho_c=-1/N) &=& \rho_c q_{d|c}[p f(AC \rightarrow AD) \nonumber \\
    & &+(1-p)f(AC \rightarrow ID)].
    \label{eq:A6}
\end{eqnarray}

Considering the expression of Fermi function with respect to the intensity of selection $\omega$, the Eq.\ref{eq: fermi} can be rewritten as
\begin{eqnarray}
    f(s_x \to s_y) &=& \frac{a_x(t)}{1+e^{[-\omega(\Pi_y-\Pi_x)]}} \nonumber \\
    &=& \frac{1}{2} a_x(t)+\frac{1}{4} \omega a_x(t) (\overline{\Pi}_y-\overline{\Pi}_x)+O(\omega^2). \nonumber \\
    \label{eq:A7}
\end{eqnarray}

We assume that every imitation event occurs in one unit of time $1/N$. Therefore, the derivative of $\rho_c$ is given by
\begin{eqnarray}
    \dot{\rho_c} &=& \left(\frac{1}{N} T^{+}(\Delta \rho_c=\frac{1}{N})-\frac{1}{N}  T^{-}(\Delta \rho_c=-\frac{1}{N})\right)N \nonumber \\
    &=& \frac{1}{4} \omega \rho_c (1-q_{c|c})[(1+p)(\overline{\Pi}_AC-\overline{\Pi}_AD)\nonumber \\
    & &+(1-p)(\overline{\Pi}_IC-\overline{\Pi}_ID)]+O(\omega^2).
    \label{eq:A8}
\end{eqnarray}

Simultaneously, a successful adoption of cooperator strategy will increase the number of CC-pairs by $1+(k-1)q_{c|d}$ and $\rho_{cc}$ increases by $\frac{1+(k-1)q_{c|d}}{k N / 2}$, thus the derivative of CC-pairs is
\begin{eqnarray}
    \dot{\rho_{cc}} &=& \left(\frac{1+(k-1)q_{c|d}}{k N / 2} T^{+}(\Delta \rho_c=\frac{1}{N}) \nonumber \right.\\
    & &\left. -\frac{1+(k-1)q_{c|d}}{k N / 2}  T^{-}(\Delta \rho_c=-\frac{1}{N})\right)N \nonumber \\
    &=& \frac{1}{k} \rho_c (1-q_{c|c}) \frac{1+(k-2)\rho_c-(k-1)q_{c|c}}{1-\rho_c}+O(\omega).\nonumber \\
    \label{eq:A9}
\end{eqnarray}

Furthermore, we can compute the derivative of $q_{c|c}$ is 
\begin{eqnarray}
    \dot{q_{c|c}} &=& \frac{\mathrm{d}}{\mathrm{d}t}\left(\frac{\rho_{cc}}{\rho_c}\right) \nonumber \\
    &=& \frac{1}{k}(1-q_{c|c})\frac{1+(k-2)\rho_c-(k-1)q_{c|c}}{1-\rho_c}+O(\omega). \nonumber \\
    \label{eq:A10}
\end{eqnarray}
Therefore, Eqs. \ref{eq:A8} and \ref{eq:A9} that describe the evolutionary dynamics of cooperation can be written as function of  $\rho_c$ and $q_{c|c}$  as

\begin{equation}
    \begin{cases}
        \dot{\rho_c}=\omega F_1(\rho_c,q_{c|c})+O(\omega^2)\\
        \dot{q_{c|c}} = F_2(\rho_c,q_{c|c})+O(\omega)
    \end{cases}.
    \label{eq:A11}
\end{equation}

Based on the dynamical equation~\ref{eq:A11} we obtained above, our next goal is to investigate the theoretical conditions for promoting cooperation in structured populations. For weak selection $\omega\rightarrow0$, the local frequency $q_{c|c}$  equilibrates much more quickly and converge to the stationary state of $F_2(\rho_c,q_{c|c})=0$  than $\rho_c$. Hence, we have
\begin{equation}
    q_{c|c}=\frac{1}{k-1}+\frac{k-2}{k-1}\rho_c.
    \label{eq:A12}
\end{equation}
Furthermore, based on Eq.\ref{eq:A12} we can obtain the following expressions
\begin{eqnarray}
    q_{d|c}&=&\frac{k-2}{k-1}(1-\rho_c), \nonumber \\
    q_{c|d}&=&\frac{k-2}{k-1}\rho_c, \\
    q_{d|d}&=&1-\frac{k-2}{k-1}\rho_c, \nonumber
    \label{eq:A13}
\end{eqnarray}
and
\begin{equation}
    \rho_{c|d} = \frac{k-2}{k-1}(1-\rho_c)\rho_c.
    \label{eq:A14}
\end{equation}

\begin{figure}
   \centering
   \includegraphics[width=7cm,keepaspectratio]{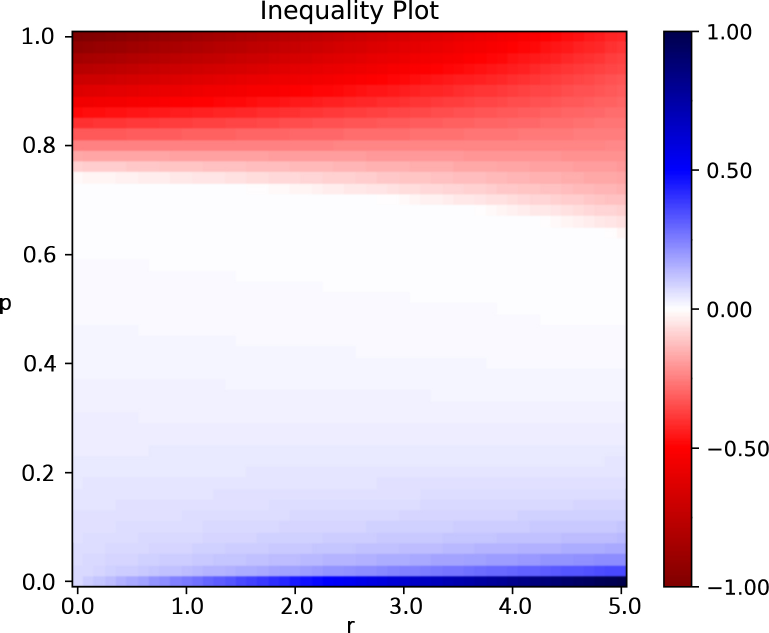}
   \caption{Numerical results of Eq.\ref{eq:A18}. the result are obtained for $k=4$, $\sigma=0.5$, $\rho_c=0.5$. Blue means that the inequation is true, red means that it is false.}
   \label{fig:A1}
\end{figure}

Accordingly, the time derivative of $\rho_c$  is
\begin{eqnarray}
    \dot{\rho_{c}} &=&\frac{k-2}{4(k-1)}\omega\rho_c(1-\rho_c)\left[(1+p)(\overline{\Pi}_{AC}-\overline{\Pi}_{AD})\right. \nonumber \\
    & &+\left.(1-p)(\overline{\Pi}_{IC}-\overline{\Pi}_{ID})\right]+O(\omega^2),
    \label{eq:A15}
\end{eqnarray}
where the differences in expected payoffs of active and inactive individuals are
\begin{eqnarray}
   & &\overline{\Pi}_{AC}-\overline{\Pi}_{AD} \nonumber \\
   &=& (1-p)^k\left(\frac{1}{k-1}+r\frac{k-p-2pk}{p(k+1)(k-1)}+r+\sigma k+1\right) \nonumber \\
   & &+\sigma k(pq_{d|d}+1-p)^k(1+pq_{d|d}) \nonumber \\
   & &+\frac{r}{k-1}\left(1-\frac{k}{p(k+1)}\right)-1,
    \label{eq:A16}
\end{eqnarray}
and
\begin{eqnarray}
    \overline{\Pi}_{IC}-\overline{\Pi}_{ID}
    &=& \sigma\left[(pq_{d|d}+1-p)^k-(pq_{d|c}+1-p)^k \right. \nonumber \\
    & &\left. +k(1-p)^{k-1}p(q_{d|c}-q_{d|d})\right].
    \label{eq:A17}
\end{eqnarray}

From Eq.\ref{eq:A15}, we can observe that evolutionary direction of the population depends on the sign of  $\overline{\Pi}_{AC}-\overline{\Pi}_{AD}$ and $\overline{\Pi}_{IC}-\overline{\Pi}_{ID}$. 
Therefore, the sufficient condition for cooperation to prevail is
\begin{equation}
    (1+p)(\overline{\Pi}_{AC}-\overline{\Pi}_{AD})
    +(1-p)(\overline{\Pi}_{IC}-\overline{\Pi}_{ID}) \geq 0.
    \label{eq:A18}
\end{equation}

For any $p\in [0,1]$ and $r\in [1,k+1]$ , we plot the colormap of in Eq.\ref{eq:A18}. As shown in Fig.\ref{fig:A1}, larger activation probability is detrimental to cooperation, and there is a negative correlation between the threshold $r$ and $p$ required to increase cooperator. By comparing the simulation results in Fig.\ref{fig:2} with the theoretical approximations, it is found that although there are some tiny deviations, the extended pair approximation method reflects role of probabilistic interaction in the spatial PGGs, especially that the transition line between all C and all D has similar trend.


\bibliography{references.bib}

\end{document}